\documentclass[a4paper,11pt]{article}
\pdfoutput=1 

\usepackage{jcappub} 

\usepackage{soul}

\usepackage[utf8]{inputenc}

\usepackage[normalem]{ulem}

\title{\boldmath 

Contributions from primordial non-Gaussianity and General Relativity to the galaxy power spectrum}

\author[a]{Rebeca Martinez-Carrillo,\note{Corresponding author.}}
\author[b]{Juan Carlos Hidalgo,}
\author[a]{Karim A. Malik,}
\author[a,c]{Alkistis Pourtsidou}

\affiliation[a]{Astronomy Unit, School of Physics and Astronomy, Queen Mary University of London,\\ Mile End Road, London, E1 4NS, United Kingdom.}

\affiliation[b]{Instituto de Ciencias F\'isicas, Universidad Nacional Aut\'onoma de M\'exico,\\  62210, Cuernavaca, Morelos, M\'exico.}

\affiliation[c]{Department of Physics and Astronomy, University of the Western Cape,\\ Cape Town 7535, South Africa}

\emailAdd{r.martinezcarrillo@qmul.ac.uk}
\emailAdd{hidalgo@icf.unam.mx}
\emailAdd{k.malik@qmul.ac.uk}
\emailAdd{a.pourtsidou@qmul.ac.uk}

\abstract{We compute the real space galaxy power spectrum, including the leading order effects of General Relativity and primordial non-Gaussianity from the $f_{\mathrm{NL}}$ and $g_{\mathrm{NL}}$ parameters. Such contributions come from the one-loop matter power spectrum terms dominant at large scales, and from the factors of the non-linear bias parameter $b_{\mathrm{NL}}$ (akin to the Newtonian $b_{\phi}$). We assess the detectability of these contributions in Stage-IV surveys. In particular, we note that specific values of the bias parameter may erase the primordial and relativistic contributions to the configuration space power spectrum.}

\keywords{galaxy power spectrum, primordial non-gaussianity, relativistic cosmology}
\arxivnumber{2107.10815}
\begin{document}
\maketitle
\flushbottom

\section{Introduction}

In recent years our  understanding of the evolution of the Universe has greatly benefited from observations. With the forthcoming Stage IV experiments such as Euclid\footnote{\url{http://euclid-ec.org}} \cite{2011arXiv1110.3193L}, the Dark Energy Spectroscopic Instrument (DESI)\footnote{\url{https://www.desi.lbl.gov/}} \cite{2016arXiv161100036D} and the Vera C. Rubin Observatory’s Legacy Survey of Space and Time (LSST)\footnote{\url{https://www.lsst.org/}} \cite{2019ApJS..242....2C}, we are expecting to obtain high-precision measurements in order to improve our understanding of the large scale structure (LSS) of the Universe. The progress achieved over the last few years demands a comprehensive description of observables from the full theory, a general relativistic description, in order to take advantage of the detail and scales these surveys aim to map. 

Most of the cosmological information is encoded in the 2-point correlation function or its Fourier space equivalent, the
power spectrum, which, among other contributions, includes  those of primordial non-Gaussianity (PNG). The possibility to constrain PNG through the power spectrum has been reported e.g.~in Refs.~\cite{2008ApJ...684L...1C,2008JCAP...08..031S,2011PhRvD..84h3509H,2012MNRAS.422.2854G,2013MNRAS.428.1116R,2015PhRvD..91d3506F,2017PhRvD..95l3513D,2018MNRAS.478.1341K,2019JCAP...09..010C,2020JCAP...11..052K,2020JCAP...12..031B,2021arXiv210706887B,2021PhRvD.104d3531P}, showing that the late-time statistics contain crucial information of the physics prevalent in the early universe. One of the main strategies to estimate PNG effects in the current matter distribution is to analyse the corresponding galaxy power spectrum, which is related to the underlying matter distribution through a set of bias parameters.
In the past years galaxy bias has been included mostly in Newtonian descriptions of the LSS \cite{2008PhRvD..77l3514D,2010PhRvD..81f3530G,2015JCAP...09..029A,2015JCAP...12..043A,2020PhRvD.102j3530E}. An extensive review, presenting some of the main studies of large scale galaxy bias can be found in Ref.~\cite{2018PhR...733....1D}.

The ever-increasing precision in the measurement of the power spectrum allows us to consider that, in addition to PNG contributions, general relativistic effects may also be observable at the large scales of the evolved matter distribution. This is due to the non-linear nature of the theory. Such effects in the galaxy clustering have been studied  for example in  Refs.~\cite{2009PhRvD..80h3514Y,2010PhRvD..82h3508Y,2011PhRvD..84d3516C,2011PhRvD..84f3505B,2012PhRvD..85b3504J,2014PhRvD..90l3507Y,2021CQGra..38f5014F,2021CQGra..38u5008F}.

In this paper we are interested in accounting for both relativistic and PNG effects in the galaxy distribution using a suitable parametrization of the galaxy bias.
Galaxy bias in a relativistic context has been previously explored including primordial non-Gaussianities, e.g.~in Refs.~\cite{2011JCAP...10..031B,2011JCAP...04..011B}, and the most suitable gauge to define the bias in, namely the comoving-synchronous gauge, has been discussed in Ref.~\cite{2012PhRvD..85d1301B}.

A derivation of the local bias at second order in cosmological perturbation theory, for Gaussian and non-Gaussian initial conditions is studied in Refs.~\cite{2019JCAP...05..020U,2019JCAP...12..048U}, where it is argued that general relativistic effects affect local clustering predominantly through the distortion of the volume element of the local patch, and that modulations of the short-modes are possible only through primordial non-Gaussianity.
At one-loop order under the weak field approximation, relativistic corrections have been included for the power spectrum at intermediate scales in Ref.~\cite{2019JCAP...07..030C}, with results extended to obtain the galaxy power spectrum using a Lagrangian bias expansion in General Relativity in Ref.~\cite{2020JCAP...07..033C}.

More recently, in Refs.~\cite{2020JCAP...11..064G,2021arXiv210608857C} it has been shown that adding all of the relativistic effect on the light propagation, avoids the infrared divergences in the galaxy power spectrum at linear order.

In this paper we extend our previous work \cite{2020JCAP...04..028M}, by not only including relativistic corrections to the power spectrum at one-loop and the input from primordial non-Gaussianity but also  introducing a bias prescription guided by the parametrization in Refs.~\cite{2006PhRvD..74j3512M, 2008PhRvD..78l3519M}, and adopt a set of two parameters, a linear $b_{\delta}$ and a non-linear parameter $b_{\mathrm{NL}}$ to compute the galaxy power spectrum in the synchronous-comoving gauge\footnote{The synchronous-comoving gauge is an adequate gauge choice to express the Lagrangian frame and the simplest to specify a galaxy bias at second order \cite{2015CQGra..32q5019B}.}. We calculate the galaxy power spectrum for a range of reasonable values of the PNG and bias parameters and compare results with the forecasted measurements from Stage-IV experiments, specifically from a $15,000$ $\mathrm{deg^2}$ and a $40,000$ $\mathrm{deg^2}$ (all-sky) galaxy survey.

In accordance to our goal, in this paper we refer to relativistic effects as those modifying the actual density contrast. That is, we consider the pure-source contributions and not the relativistic effects on the light propagation that influence the observed power spectrum; a task that we leave for future work.

The paper is structured as follows: in section~\ref{Relativistic-corrections-model} we briefly review previous work and present the expressions for the relativistic density contrast that we used in the computation of the galaxy power spectrum. In section~\ref{galaxy-galaxy-section} we introduce the bias model and the mathematical expression of the galaxy power spectrum. In section~\ref{Results-section} we present our results in a series of plots of the galaxy power spectrum with a range of values for the PNG parameters and for the non-linear bias.
 The spectra include the observational uncertainties from Stage-IV galaxy surveys. In this section we note  that a specific combination of parameter values could cancel the PNG and relativistic signatures at large scales. Finally in section~\ref{Discussion-section} we discuss our results.

Throughout this paper we use the conformal time $\eta$, and denote derivatives with respect to $\eta$ with a prime. Greek indices $\mu$, $\nu$, range from 0 to 3, lower case Latin indices, $i$, $j$, and $k$, range from 1 to 3.

\section{Relativistic density contrast}
\label{Relativistic-corrections-model}
In this section we present the basic definitions for our calculation, the expressions for the relativistic density contrast. These definitions were presented in previous works and the details of the calculation can be found in e.g.~Refs.  \cite{2020JCAP...04..028M,2018JCAP...06..016G,2014ApJ...785....2B,2014ApJ...794L..11B}. We briefly summarise how these are obtained to facilitate the reader's comprehension of further sections.

The starting point of this work is the perturbed Friedmann-Robertson-Walker line element in synchronous-comoving gauge \cite{2009PhR...475....1M,2008CQGra..25s3001M}, 
\begin{equation}
ds^2=a^2(\eta)[-d\eta^2+\gamma_{ij}dx^idx^j] \, ,
\end{equation}
where $a$ is the scale factor and $\gamma_{ij}$ the perturbed spatial metric.
The matter content considered is a pressureless and irrotational fluid. The density field $\rho$ is given by 
\begin{equation}
\rho(\mathbf x,\eta)=\bar\rho(\eta)+\delta\rho(\mathbf x,\eta)=\bar\rho(\eta)(1+\delta(\mathbf x,\eta)) \, ,
\end{equation}
where $\bar\rho(\eta)$ is the density in the background, $\delta\rho(\mathbf x,\eta)$ is a small perturbation and $\delta(\mathbf x,\eta)$ is the density contrast. 

The evolution of the density contrast $\delta(\mathbf x,\eta)$ is given by the continuity equation 
\begin{equation}
\delta'+(1+\delta)\vartheta=0 \, ,
\label{continuity}
\end{equation}
where $\vartheta$ is the trace of the deformation tensor $\vartheta_\nu^{\mu}$, that is $\vartheta=\vartheta_\alpha^\alpha$, and the deformation tensor is  defined by 
\begin{equation}
\vartheta_\nu^{\mu}\equiv au^{\mu}_{~;\nu}-\mathcal{H}\delta_\nu^{\mu}\,,
\end{equation}
where $u^\mu$ is the fluid 4-velocity, $\mathcal{H}$ the Hubble parameter in conformal time.
The evolution for $\vartheta$ is given by the Raychaudhuri equation \cite{2014ApJ...785....2B,Meures}
\begin{equation}
\vartheta'+\mathcal{H}\vartheta+\vartheta_j^i\vartheta_i^j+4\pi G a^2 \bar\rho\delta=0.
\label{Ray}
\end{equation}
The energy constraint is given by 
\begin{equation}
\vartheta^2-\vartheta_j^i\vartheta_i^j+4\mathcal{H}\vartheta+{}^{3}R=16\pi Ga^2\bar\rho\delta \, ,
\label{energycons}
\end{equation}
where ${}^{3}R$ is the spatial Ricci scalar of the spatial metric $\gamma_{ij}$. 

Finding a solution for the density contrast $\delta$ can be achieved by using either perturbation theory or a gradient expansion; for brevity we present the gradient approach only. In this approach the spatial part of the metric is written in terms of a conformal factor as \cite{1990PhRvD..42.3936S,2005JCAP...05..004L}
\begin{equation}
g_{ij}=a^2\gamma_{ij}=a^2e^{2\zeta}\check\gamma_{ij}.
\end{equation}
Using this approximation, spatial gradients are small compared to time derivatives. In addition, we find that
\begin{equation}
\delta\sim\vartheta\sim{}^3R\sim\nabla^2.
\end{equation}
This means that on large scales, the conformal metric can be approximated as $\check\gamma_{ij}\simeq\delta_{ij}$. The result of this simplified metric is having the Ricci scalar as a function of the curvature perturbation $\zeta$ only \cite{2018JCAP...06..016G,2014ApJ...794L..11B}

\begin{equation}
{}^{3}R=-4\nabla^2\zeta+\sum_{m=0}^{\infty}\frac{\left(-2\right)^{m+1}}{(m+1)!}\left[(m+1)(\nabla\zeta)^2-4\zeta\nabla^2\zeta\right]\zeta^m \, ,
\label{Ricciexpansion}
\end{equation}
where $\zeta$ is expanded up to third order as
\begin{equation}
\zeta=\zeta^{(1)}+\frac{3}{5}f_{\mathrm{NL}}\zeta^{(1)2}+\frac{9}{25}g_{\mathrm{NL}}\zeta^{(1)3}\,,
\end{equation}
with $f_{\mathrm{NL}}$ and $g_{\mathrm{NL}}$ being the usual local non-Gaussianity parameters \cite{2010CQGra..27l4002W}.

The general solution for the density contrast in Einstein de-Sitter is of the form 
 
 \begin{equation}
 \delta(\eta,\mathbf x)=C(\mathbf x)D_{+}(\eta) \, ,
 \label{gralsolu}
 \end{equation}
 where $C(\mathbf x)$ is given by \cite{2014ApJ...785....2B}
 \begin{equation}
C(\mathbf x)=\frac{{}^{3}R}{10\mathcal{H}_{IN}^2D_{+IN}} \, ,
\end{equation}
 where $D_{+}$ is a constant -- the growth factor evaluated at some time early in the matter dominated era, denoted by subscript ``$IN$". 

Using Eq.~\eqref{Ricciexpansion}, the relativistic density contrast in Eq.~\eqref{gralsolu} up to third order is given by
\begin{equation}
\delta=\delta^{(1)}+\frac{1}{2}\delta^{(2)}+\frac{1}{6}\delta^{(3)},
\label{reldensity}
\end{equation}
where \cite{2020JCAP...04..028M}
\begin{align}
\label{delta1}
\delta^{(1)}=&\frac{D_{+}(\eta)}{10\mathcal{H}_{IN}^2D_{+IN}}\left(-4\nabla^2\zeta^{(1)}\right),\\
\label{delta2}
\delta^{(2)}=&\frac{D_{+}(\eta)}{10\mathcal{H}^2_{IN}D_{+IN}}\frac{24}{5}\Bigg[-(\nabla\zeta^{(1)})^2\bigg(\frac{5}{12}+f_{\mathrm {NL}}\bigg)+\zeta^{(1)}\nabla^2\zeta^{(1)}\bigg(\frac{5}{3}-f_{\mathrm {NL}}\bigg)\Bigg],\\
\label{delta3}
\delta^{(3)}=&\frac{D_{+}(\eta)}{10\mathcal{H}^2_{IN}D_{+IN}}\frac{108}{25}\Bigg[2\zeta^{(1)}(\nabla\zeta^{(1)})^2\bigg(-g_{\mathrm {NL}}+\frac{5}{9}f_{\mathrm {NL}}+\frac{25}{54}\bigg)\\
\nonumber
+&\zeta^{(1)2}\nabla^2\zeta^{(1)}\bigg(-g_{\mathrm {NL}}+\frac{10}{3}f_{\mathrm{ NL}}-\frac{50}{27}\bigg)\Bigg],\\
\nonumber
\end{align}
and the growth factor in Einstein-de Sitter is 
\begin{equation}
D_{+}=\frac{D_{+IN}\mathcal{H}_{IN}^2}{\mathcal{H}^2},
\end{equation}
with $D_{+IN}=1$ and $\mathcal{H}_{IN}=\mathcal{H}_0$, where $\mathcal{H}_0$ is the conformal Hubble parameter at present time. 

\section{Galaxy power spectrum}
\label{galaxy-galaxy-section}
To compute the galaxy power spectrum we follow the renormalized perturbative bias model of Refs.~\cite{2006PhRvD..74j3512M, 2008PhRvD..78l3519M}. This approach assumes the Taylor expansion of a general function of the density contrast, given by
\begin{equation}
\delta_g=c_\delta\delta+\frac{1}{2}c_{\delta^2}(\delta^2-\sigma^2)+\frac{1}{3!}c_{\delta^3}\delta^3+ \epsilon+\mathcal{O}(\delta^4),
\label{MCdensity}
\end{equation}
where $\sigma^2=\langle\delta^2\rangle$ is the variance of $\delta$, the coefficients of the Taylor expansion $c_{\delta^n}$ constitute the bias parameters, while $\epsilon$ is a random noise variable that allows for stochasticity. This prescription was originally defined for matter density in the Eulerian frame, though we adopt it for the Lagrangian density expressed earlier. Our parameters should thus be interpreted as Lagrangian variables\footnote{The Lagrangian bias parameters should coincide with the Eulerian set at large scales, because the coordinate change to the Eulerian frame alters quantities at small scales only \cite{2014ApJ...785....2B,2015CQGra..32q5019B}.}.

If we consider the density contrast in perturbative orders up to the leading non-linear contribution to the power spectrum, then Eq.~\eqref{MCdensity} is expanded as
\begin{equation}
\delta_g=c_\delta\left(\delta^{(1)}+\frac{\delta^{(2)}}{2}+\frac{\delta^{(3)}}{6}\right)+\frac{c_{\delta^2}}{2}\left(\delta^{(1)2}+\delta^{(1)}\delta^{(2)}\right)+...,
\label{galaxy-density}
\end{equation}
where $\delta^{(1)}$, $\delta^{(2)}$ and $\delta^{(3)}$ are defined in Eqs.~\eqref{delta1}, \eqref{delta2} and \eqref{delta3} respectively.

The galaxy power spectrum is defined as
\begin{equation}
\langle\delta_g(k,\eta)\delta_g(k',\eta)\rangle=(2\pi)^3P_{gg}(k,\eta)\delta_D(k+k'),
\label{galaxypower}
\end{equation}
where $k$ is the comoving wavenumber in Fourier space. The leading order non-linear corrections to the galaxy power spectrum are of order $\delta^{(4)}$ only. Using Eq.~\eqref{galaxy-density} and Eq.~\eqref{galaxypower}, the galaxy power spectrum $P_{gg}(k,\eta)$ is given by
\begin{equation}
\label{eq:g-spectrum1}
P_{gg}(k,\eta)=(c_\delta)^2\left[P_L(k,\eta)+2P_R^{(1,3)}(k,\eta)+P_R^{(2,2)}(k,\eta)\right]+(2c_{\delta}c_{\delta^2})\left[P_{R1}(k,\eta)+P_{R2}(k,\eta)\right],
\end{equation}
where $P_L(k,\eta)$ represents the linear power spectrum, $P_R^{(1,3)}(k,\eta)$ and $P_R^{(2,2)}(k,\eta)$ are the relativistic contributions to the  one-loop matter power spectrum (previously presented in Ref.~\cite{2020JCAP...04..028M}), and are given by 
\begin{align}
P_R^{(1,3)}(k,\eta)=&\frac{k^3}{(2\pi)^2}P_L(k,\eta)\int_{0}^{\infty} drP_L(kr,\eta)\Bigg\{81
\left(\frac{\mathcal{H}^4}{k^4}\right)
\bigg[\Big(-g_{\mathrm{NL}}+\frac{5}{9}f_{\mathrm{NL}}+\frac{25}{54}\Big)\\\nonumber
&+\frac{1+2r^2}{6r^2}\Big(g_{\mathrm{NL}}-\frac{10}{3}f_{\mathrm{NL}}+\frac{50}{27}\Big) \bigg]\Bigg\},\\
P_R^{(2,2)}(k,\eta)=&\frac{k^3}{2\pi^2}\int_{0}^{\infty}dr  P_L(kr,\eta)\int_{-1}^{1}dxP_L(k\sqrt{1+r^2-2rx},\eta)\\
  &\times\Bigg\{\left(\frac{{\mathcal{H}^4}}{k^4}\right)\bigg[\frac{6f_{\mathrm {NL}}-10-25r^2+25rx}{4r(1-2rx+r^2)}\bigg]^2\Bigg\}.
\nonumber
\end{align}
The  contributions $P_{R1}(k,\eta)$ and $P_{R2}(k,\eta)$, which are not present in the one-loop matter spectrum, are given by
\begin{align}
\nonumber
P_{R1}(k,\eta)=&\frac{k^3}{(2\pi)^2}P_L(k,\eta)\int_{0}^{\infty} drP_L(kr,\eta)\\
&\times\int_{-1}^{1} dx\left(\frac{\mathcal{H}^2}{k^2}\right)\left[3f_{\mathrm{NL}}(1+r^2+2rx)+\frac{5}{6}(3rx-6+6r^2)\right],\\
\nonumber
P_{R2}(k,\eta)=&\frac{k^3}{(2\pi)^2}\int_{0}^{\infty} drP_L(kr,\eta)\int_{-1}^{1} dxP_L(k\sqrt{1+r^2-2rx},\eta)\\
&\times\left(\frac{\mathcal{H}^2}{k^2}\right)\left[\frac{6f_{\mathrm{NL}}-10-25r^2+25rx}{4(1-2rx+r^2)}\right].
\end{align}

We now re-express the bias factors in terms of the linear and non-linear bias parameters for $P_{gg}(k,\eta)$ in terms of the $c_{\delta}$ and $c_{\delta^2}$. At leading order the correspondence is straightforward and is reduced to parameters
\begin{align}
b_{\delta}=&c_{\delta}\,,\quad
b_{\mathrm{NL}}=2c_{\delta^2}\,,
\end{align}
and we get
\begin{equation}
\label{eq:g-spectrum2}
P_{gg}(k,\eta)=(b_\delta)^2\left(P_L(k,\eta)+2P_R^{(1,3)}(k,\eta)+P_R^{(2,2)}(k,\eta)\right)+(b_{\delta}b_{\mathrm{NL}})\Big(P_{R1}(k,\eta)+P_{R2}(k,
\eta)\Big).
\end{equation}

The  bias parameter $b_{\mathrm{NL}}$ corresponds to $b_{\phi}$ introduced in Ref.~\cite{2008PhRvD..78l3519M}, the latter accounts for the dominant contributions from primordial non-Gaussianity. We use a different label to emphasise that $b_{\mathrm{NL}}$ includes PNG as well as relativistic terms. Note, that the terms multiplied by $b_{\mathrm{NL}}$ include those of the \textit{scale-dependent bias}, which is the focus of previous studies (see e.g.~Refs.~\cite{2008PhRvD..77l3514D,2012PhRvD..85d1301B}). 

As can be seen from Eq.~\eqref{eq:g-spectrum2}, with the parametrization chosen above, the one-loop matter power spectrum is included in the terms multiplied by the square of the linear bias, $(b_\delta)^2$. This is consistent with the fact that the relativistic solutions are part of the linear correspondence between matter density and spatial curvature (the lowest order solution in the gradient expansion).  This characteristic feature of General Relativity and PNG plays a crucial role in the bias parameter fitting as we will see below.

\section{Results}
\label{Results-section}

In this section we present the (source) galaxy power spectrum for a set of values of interest of the galaxy bias parameters. This is best appreciated in plots of the galaxy power spectrum. All our calculations were performed numerically in Python, using as an input a linear power spectrum generated with the \texttt{CLASS} Boltzmann solver \cite{2011arXiv1104.2932L,2011JCAP...07..034B}, for realisations of the $\Lambda$CDM cosmology, taking parameter values from  the Planck collaboration results \cite{2016A&A...594A..13P}.

In all our plots of the galaxy power spectrum, the value for the linear bias parameter is fixed to $b_{\delta}=1.41$ in a survey bin with mean redshift $z=1$, following the parametrisation \cite{2020A&A...642A.191E}

\begin{equation}
b_\delta=\sqrt{1+z} \, .
\label{def_lin_bias}
\end{equation}

\begin{figure}[tbp]
\centering
\includegraphics[width=140mm]{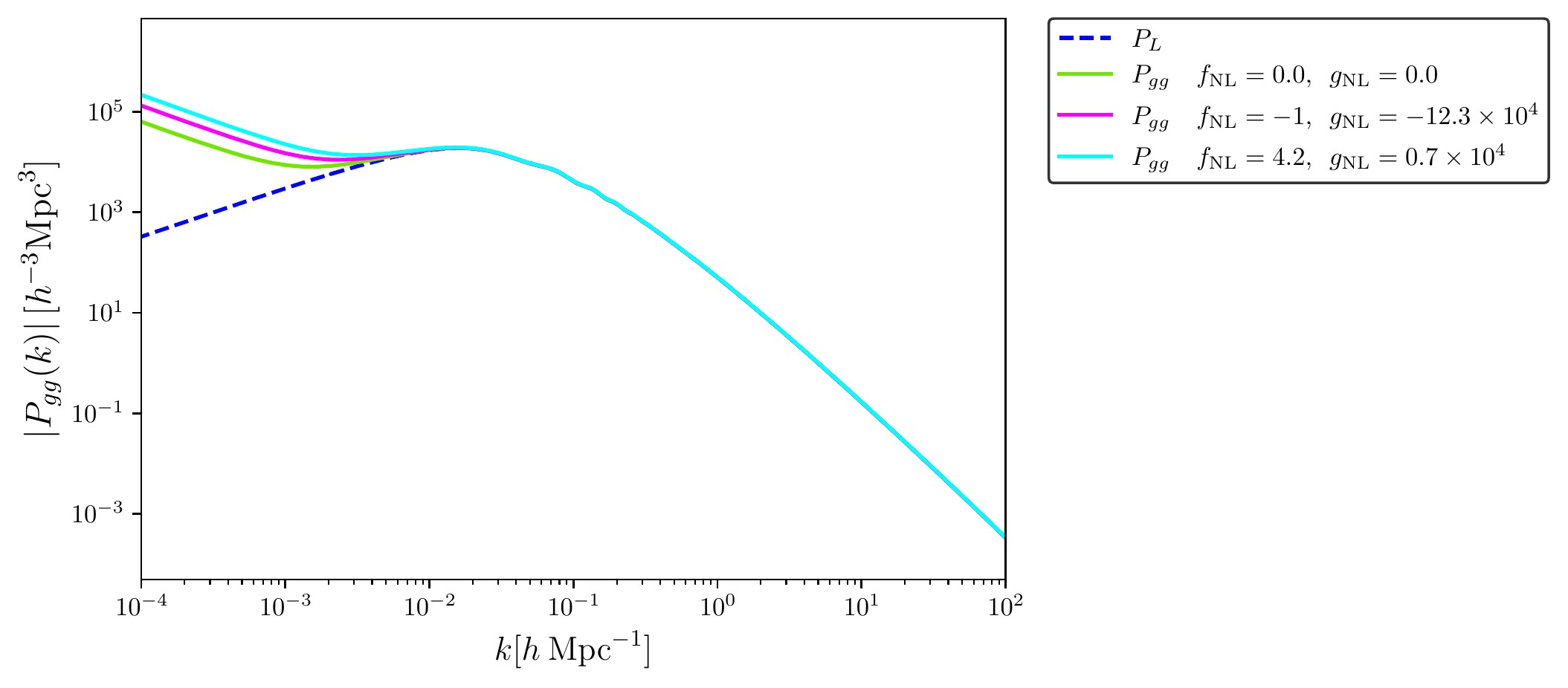}
\caption[Galaxy power spectrum in a wide scale range, at redshift $z=1$, $b_\delta=1.41$ and $b_{\mathrm{NL}}=0.2$, for different limiting  values of $f_{\mathrm {NL}}$ and $g_{\mathrm{NL}}$.]{Galaxy power spectrum in a wide scale range, at redshift $z=1$, $b_\delta=1.41$ and $b_{\mathrm{NL}}=0.2$, for different limiting  values of $f_{\mathrm {NL}}$ and $g_{\mathrm{NL}}$ reported by Planck \cite{2020A&A...641A...9P}. No divergence is observed in any of the cases at small scales (large $k$-modes). The divergences at the other end are discussed in section~\ref{sec:renorma}.}
\label{allscales}
\end{figure}

First,  we show that there is no divergence of the non-linear contributions at small scales. We plot in Fig.~\ref{allscales} the galaxy power spectrum of Eq.~\eqref{eq:g-spectrum2} at $z=1$. This includes the non-Gaussianity contributions with the limiting values of parameters $f_{\mathrm {NL}}$ and $g_{\mathrm {NL}}$ reported by the Planck collaboration \cite{2020A&A...641A...9P} (the value of $b_{\mathrm{NL}}$ for this figure is arbitrary within the perturbative expansion hierarchy).

Let us stress at this point, a technical but important issue, relevant for all plots in this paper. We show throughout the minimum value of $f_{\mathrm{NL}}=-1$, since smaller values yield dominant (negative) contributions to the matter power spectrum on large scales. In that sense, in the non-linear contributions to the matter power spectrum  we discarded the possibility of using  values of $g_{\mathrm{NL}}>7$ \cite{2020JCAP...04..028M}. We note, however, that such a restriction is not necessary in the galaxy power spectrum in general. 
This is because the two new contributions $P_{R1}(k,\eta)$ and $P_{R2}(k,\eta)$ may balance the original one-loop terms as dictated by the values of the $b_{\mathrm{NL}}$ parameter as exemplified below.

\begin{figure}[htbp]
\centering
\includegraphics[width=130mm]{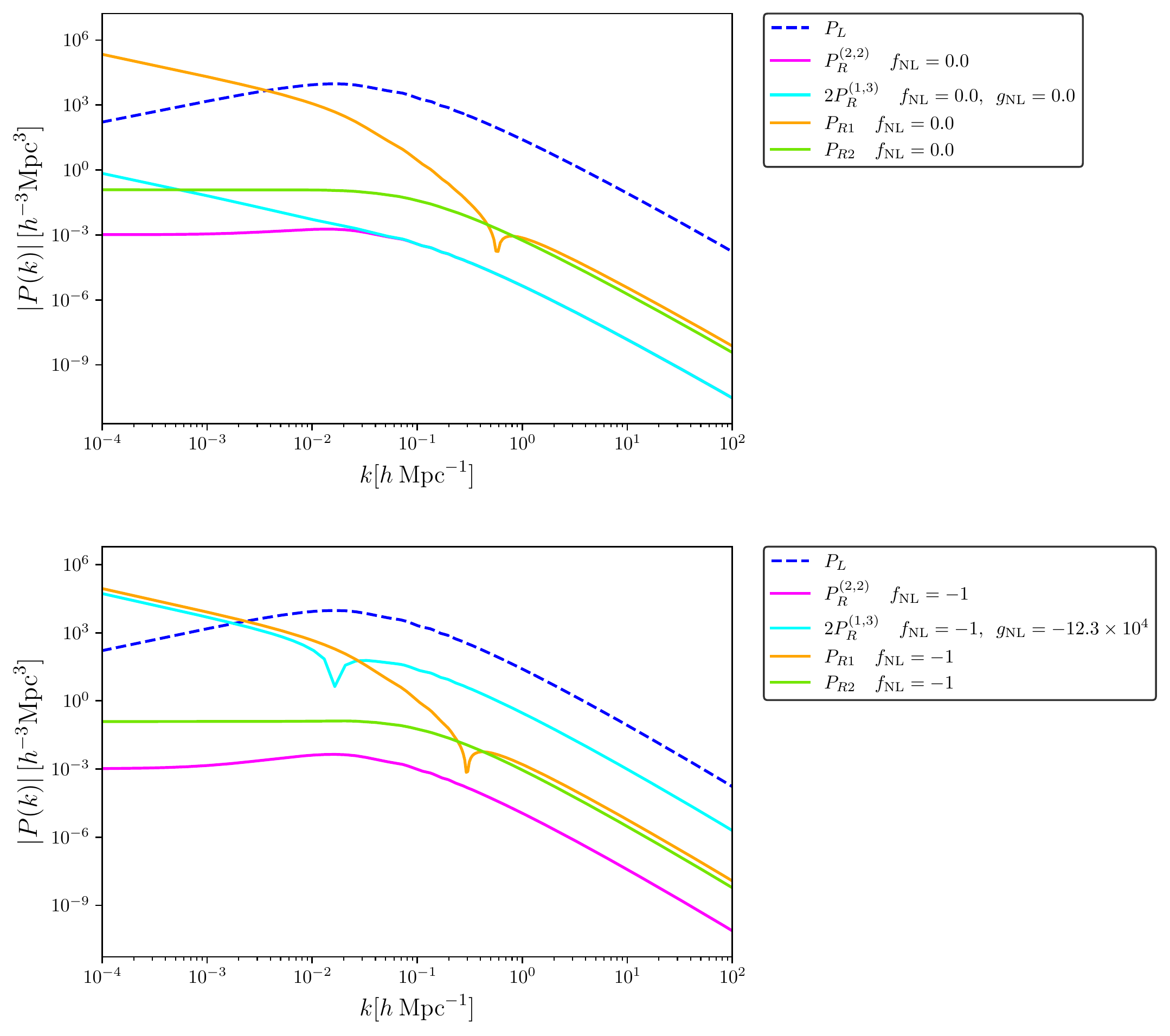} 
  
  \caption[Contributions from each term to the galaxy power spectrum in Eq.~\eqref{eq:g-spectrum2} at redshift $z=1$, for Gaussian initial conditions and for the limiting values of $f_{\mathrm{NL}}$ and $g_{\mathrm{NL}}$.]{Contributions from each term to the galaxy power spectrum in Eq.~\eqref{eq:g-spectrum2} at redshift $z=1$, for Gaussian initial conditions (upper plot), and for the limiting values of $f_{\mathrm{NL}}$ and $g_{\mathrm{NL}}$ reported by Planck \cite{2020A&A...641A...9P} (lower plot).
  }
  \label{separate-contributions}
\end{figure}

In Fig.~\ref{separate-contributions} we plot separately the contributions to the galaxy power spectrum by each term in Eq.~\eqref{eq:g-spectrum2} (setting both bias parameters to unity). The dominant contribution at large scales comes from $P_{R1}(k,\eta)$, an additional term to the one-loop matter spectrum,  followed by $P^{(1,3)}_R(k,\eta)$, which is the only term with $g_{\mathrm{NL}}$ dependence.

 Using Eq.~\eqref{def_lin_bias} for $b_\delta$, in the rest of this section we present the galaxy power spectrum for values for $b_{\mathrm{NL}}$ following two criteria: First we compute values for which the spectrum deviates from the linear prescription beyond the uncertainty in current and future surveys, and thus show observable relativistic or PNG contributions, and subsequently we compute values which cancel the divergent part of the relativistic contribution at large scales, thus showing observable features at large scales only in the presence of primordial non-Gaussianity.  

\subsection{Viable bias parameter values}
\label{sec:results1}

\begin{figure}[htbp]
\centering
\includegraphics[width=140mm]{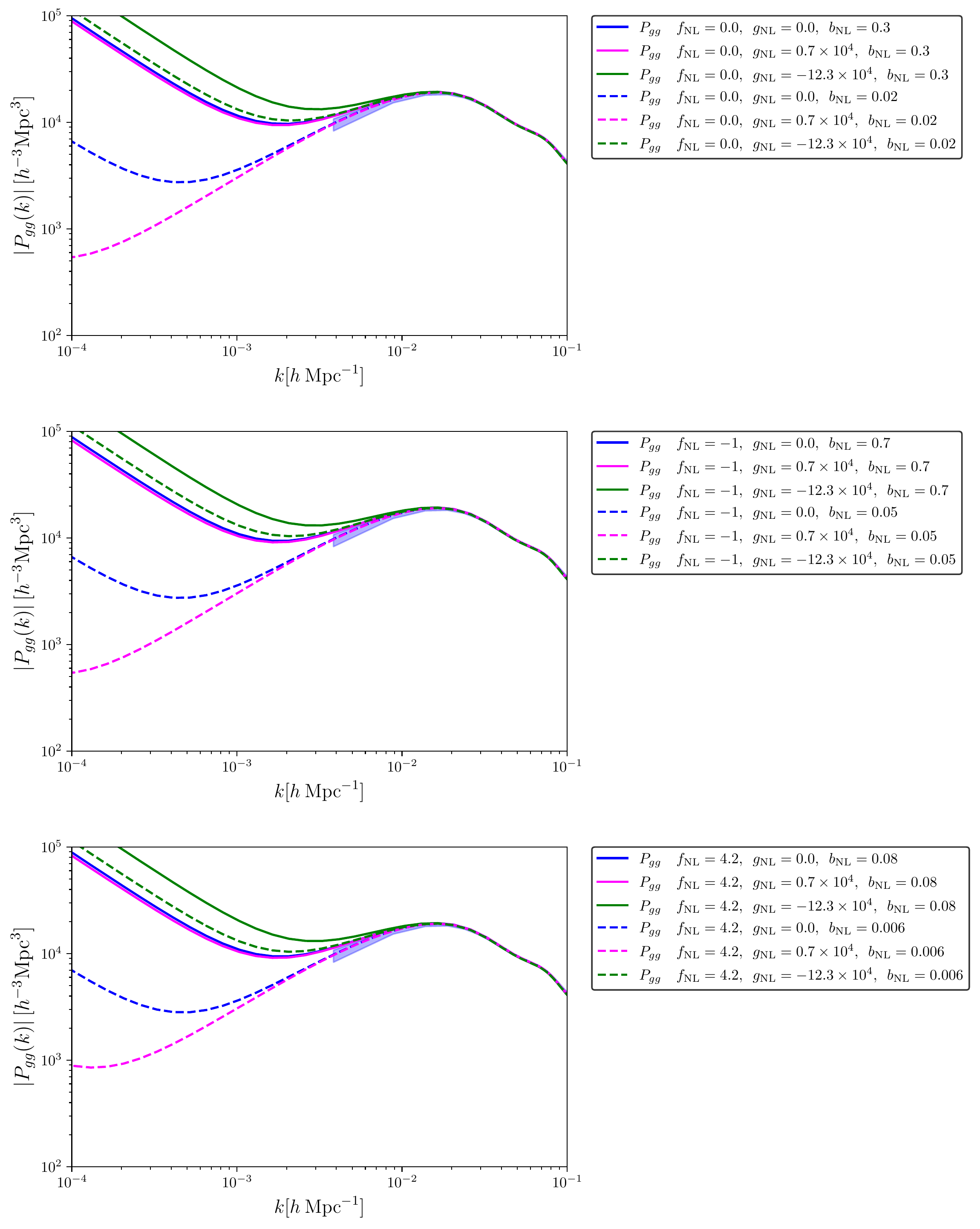}
 \caption[Galaxy power spectrum at redshift $z=1$, with $b_\delta=1.41$, for the limiting values of $f_{\mathrm{NL}}$ and $g_{\mathrm{NL}}$. The blue shaded area corresponds to the forecasted $1\sigma$ uncertainties of a cosmic variance limited survey of $15,000$ $\mathrm{deg}^2$.]{Galaxy power spectrum at redshift $z=1$, with $b_\delta=1.41$ (a choice justified in the text), for the limiting values of $f_{\mathrm{NL}}$ (zero for the top plot, $-1$ for the middle plot, and 4.2 for the bottom) each with limiting  $g_{\mathrm{NL}}$ values reported by Planck \cite{2020A&A...641A...9P}. The blue shaded area corresponds to the forecasted $1\sigma$ uncertainties of a cosmic variance limited survey of $15,000$ $\mathrm{deg}^2$. See text for more details.}
  \label{Survey15000more}
\end{figure}

\begin{figure}[htbp]
  \centering
    \includegraphics[width=140mm]{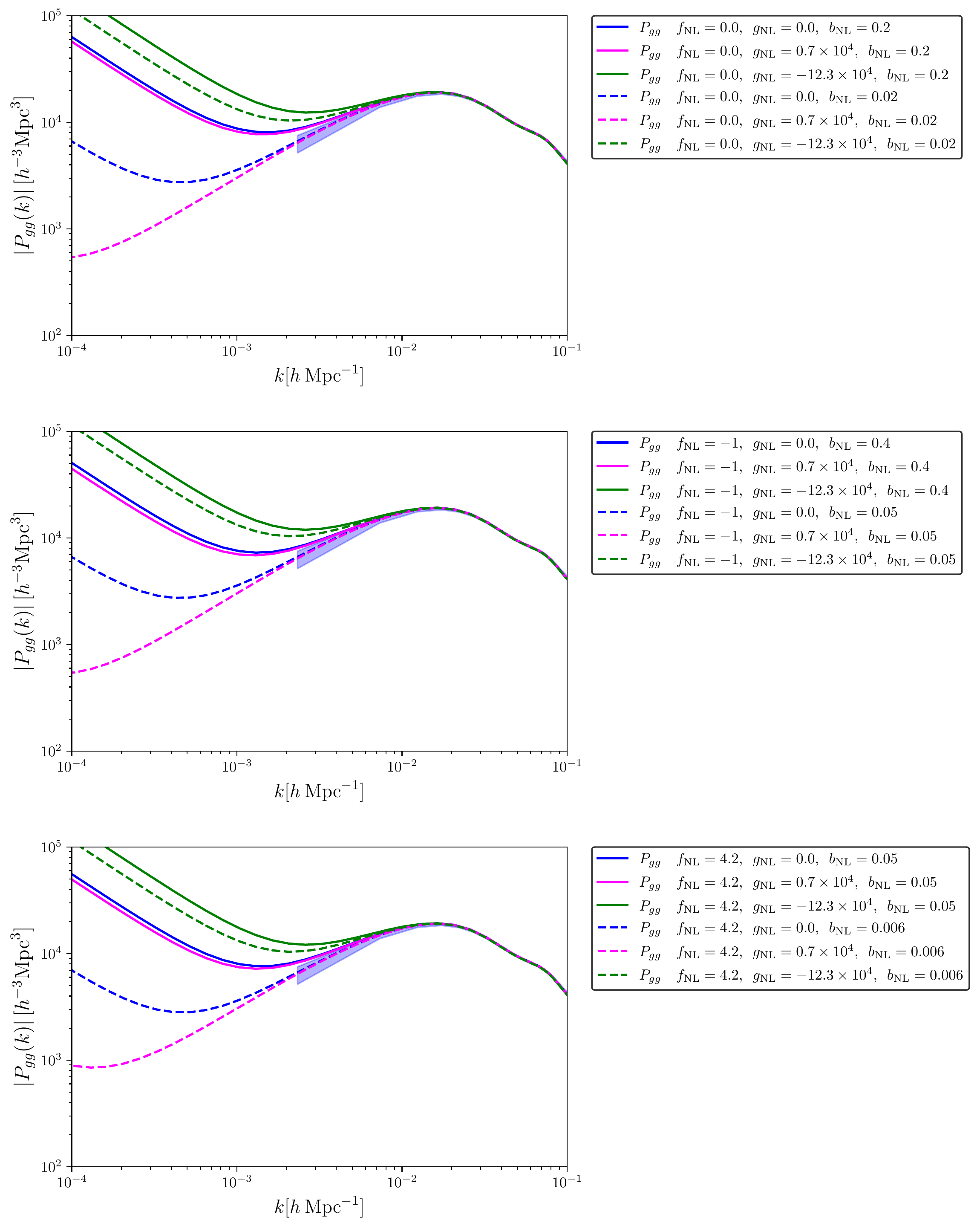}
  \caption[Galaxy power spectrum at redshift $z=1$, with $b_\delta=1.41$, for the limiting values of $f_{\mathrm{NL}}$ and $g_{\mathrm{NL}}$.The blue shaded area corresponds to the forecasted $1\sigma$ uncertainties of a cosmic variance limited survey of $40,000$ $\mathrm{deg}^2$. ]{Galaxy power spectrum at redshift $z=1$, with $b_\delta=1.41$ (a choice justified in the text), for the limiting values of $f_{\mathrm{NL}}$ (zero for the top plot, $-1$ for the middle plot, and 4.2 for the bottom) each with limiting  $g_{\mathrm{NL}}$ values reported by Planck \cite{2020A&A...641A...9P}. The blue shaded area corresponds to the forecasted $1\sigma$ uncertainties of a cosmic variance limited survey of $40,000$ $\mathrm{deg}^2$. See text for more details.}
  \label{Survey40000more}
\end{figure}

In Fig.~\ref{Survey15000more} and Fig.~\ref{Survey40000more} we present the galaxy spectrum at redshift $z=1$ for combinations of the limiting values of $f_{\mathrm {NL}}$ and $g_{\mathrm {NL}}$. In each plot we display a shaded area, corresponding to the predicted $1\sigma$ measurement errors for a survey of $15,000$ $\mathrm{deg^2}$ (Euclid-like, Fig.~\ref{Survey15000more}), and a survey of $40,000$ $\mathrm{deg^2}$ in Fig.~\ref{Survey40000more}.

Our forecasts are suitable to fit the source galaxy power spectrum and assume idealised cosmic variance limited surveys, i.e., with negligible shot noise \cite{1997PhRvL..79.3806T}. Other specifications are the redshift bin width $\Delta z=1.0$ at a central redshift $z=1$, and the largest measured scale $k_{\mathrm {min}}\simeq2\pi/V_{\mathrm {bin}}^{1/3}=0.001h\mathrm{Mpc}^{-1}$ for a $40,000 \, {\mathrm{deg}}^2$ survey and $k_{\mathrm {min}}=0.002h\mathrm{Mpc}^{-1}$ for a $15,000 \, {\mathrm{deg}}^2$ survey. We note that shot noise contributions and, most importantly, large scale systematic effects (see e.g. \cite{2021arXiv210613725M}) are expected to increase the error budget.
We fix the value of $b_{\mathrm{NL}}$ to show in solid lines the minimum value required to have a galaxy power spectrum that could be distinguished in forthcoming surveys considering the forecasted $1\sigma$ measurement errors.

Additionally, we show in dashed lines the smallest $b_{\mathrm{NL}}$ values that result in a well behaved galaxy power spectrum. As already stated, the fiducial value for $b_\delta$ is chosen to be $b_{\delta}=1.41$ at a mean redshift $z=1$.

\subsection{Bias values suppressing large-scale divergences}
\label{sec:renorma}

As mentioned earlier, recent works have argued that a divergent behaviour on large scales of the galaxy power spectrum due to the relativistic corrections is nonphysical at linear order \cite{2020JCAP...11..064G,2021arXiv210608857C}. 

\begin{figure}[htbp]
\centering
\includegraphics[width=140mm]{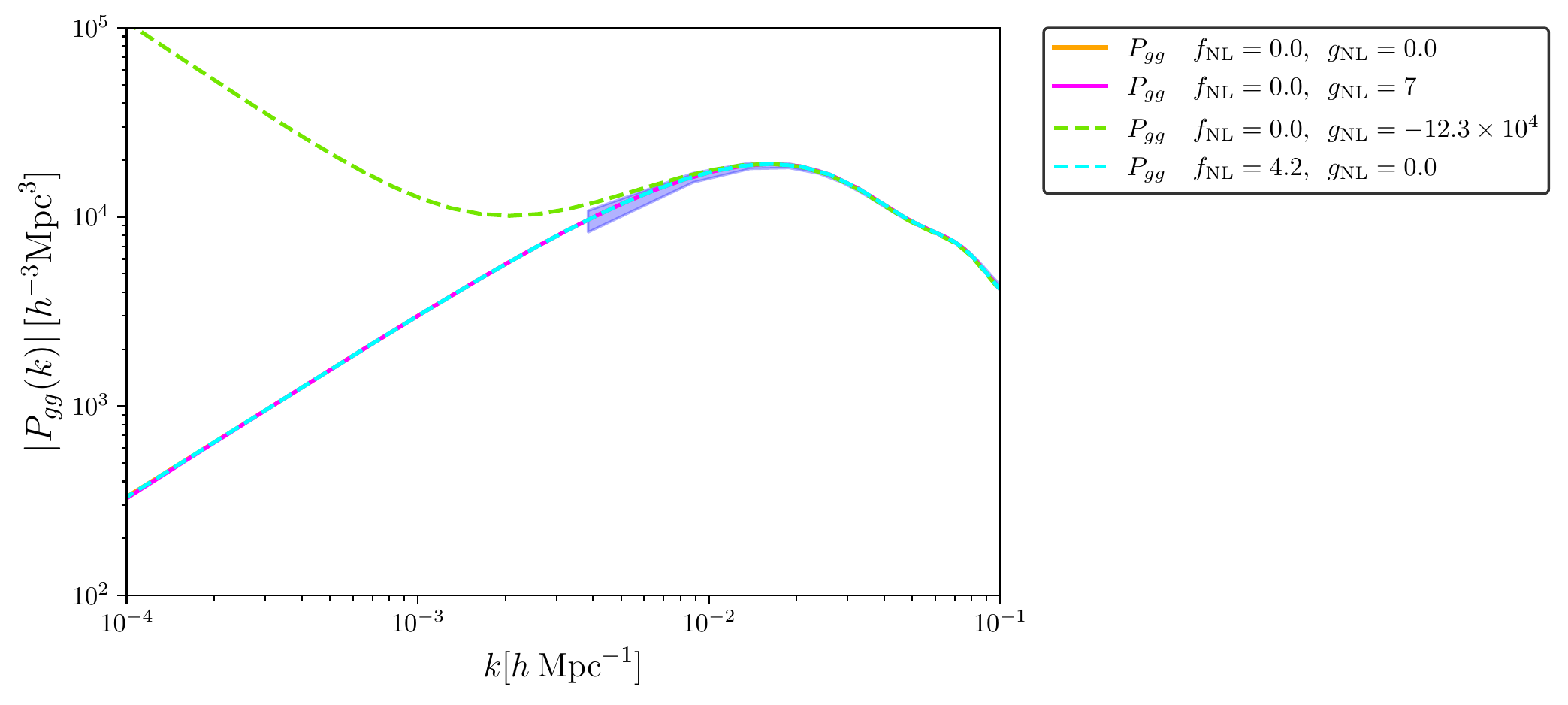}
\caption[Galaxy power spectrum, at redshift $z=1$, with the specific choice of $b_\delta=1.41$, and $b_{\mathrm{NL}}=2.44\times10^{-6}$.]{Galaxy power spectrum, at redshift $z=1$, with the specific choice of $b_\delta=1.41$, and $b_{\mathrm{NL}}=2.44\times10^{-6}$. The lines show limiting values of $f_{\mathrm {NL}}$ and $g_{\mathrm{NL}}$ as reported by Planck \cite{2020A&A...641A...9P}. The shaded area is as described in Fig.~\ref{Survey15000more}. Note that only the largest $g_{\mathrm{NL}}$ value yields a distinguishable departure from the linear spectrum.}
\label{renormalized}
\end{figure}

Here we show through an explicit example that it is possible to  suppress the divergences with   a specific choice of the new $b_{\mathrm{NL}}$ parameter. This should not be regarded as a \textit{renormalization} method but we note that the specific value $b_{\rm NL} = 2.44 \times 10^{-6}$ cancels the relativistic effects of the dominant terms $P_{R1}(k,\eta)$ and $P_{R}^{(1,3)}(k,\eta)$ at large scales, in the absence of PNG \footnote{For this bias value, the maximum value of $g_{\mathrm{NL}}$ that keeps the perturbation theory hierarchy is of the order of $g_{\mathrm{NL}}\sim 7$, since the use of higher values, even though are allowed by Planck, leads to negative contributions to the power spectrum.}. The result of this  parameter value is presented in Fig.~\ref{renormalized}. Note that when adding primordial non-Gaussianity contributions the only case that departs significantly from the linear spectrum is when $g_{\mathrm{NL}}$ takes the minimum value allowed by the Planck results.

\section{Discussion}
\label{Discussion-section}

In this paper we have computed the non-linear (source) galaxy power spectrum including the leading order relativistic and primordial non-Gaussianity contributions to the density contrast, modulated through a set of bias parameters that follow closely the standard prescriptions. 

Besides identifying the well known scale-dependent bias feature, we have non-linear contributions from the relativistic treatment of the matter density. The main result is the galaxy power spectrum including one-loop corrections, expressed in Eq.~\eqref{eq:g-spectrum2}, with the different elements plotted and discussed in detail in section~\ref{Results-section}. In order to assess the relevance of such contributions, our plots include the predicted $1\sigma$ uncertainties of planned Stage-IV galaxy surveys, thus showing which parameter values may be discriminated in future galaxy catalogues. 
An important first result is that, even in the absence of PNG, some values of the non-linear bias parameter could be discriminated by future surveys ($b_{\mathrm{NL}} \gtrsim 0.3$), which, conversely means that relativistic contributions could be detected in the galaxy power spectrum. On the other hand, such signal is degenerate with that of a large $f_{\mathrm{NL}}$ since the scale-dependence is identical for relativistic and PNG terms at large scales (see e.g.~Fig.~\ref{separate-contributions}). 

The so-called universality relations between bias parameters would fix values of $b_{\phi}$ allowing to debias contributions of PNG parameters in the Newtonian formalism \cite{2020JCAP...12..031B,2021arXiv210706887B,2021arXiv210615604M}. However, we find that the corresponding values for $b_{\mathrm{NL}}$ from the universality relation are larger than all the examples  here of section~\ref{sec:results1}.
This calls for numerical simulations of galaxy formation and complementary probes of non-Gaussianity in the galaxy or lensing maps, in order to reanalyse the correspondence between bias parameters and also to disentangle relativistic and PNG contributions. 

We note that a specific value of the non-linear bias parameter $b_{\mathrm{NL}}$ can be chosen in order to cancel the divergences of the different contributions at the largest scales, as we show in section~\ref{sec:renorma}. This does not represent an effective renormalization of the general relativistic corrections to the galaxy power spectrum. Instead, it is only a warning over the parameter values that might hide the relativistic and some of the PNG effects. In this case, and as shown in Fig.~\ref{renormalized}, the non-Gaussianity contributions $g_{\mathrm{NL}}$ can be distinguished from relativistic corrections. The divergent behaviour of relativistic contributions in the non-linear galaxy density is also present in other bias parameter choices \cite{2019JCAP...05..020U,2019JCAP...12..048U}. As in our case, the factors of more than one bias parameter show divergences of order $P_L / k^2$. Then situations as the one presented in section~\ref{sec:renorma} would also arise there.

It is important to note that an extreme value of the $g_{\mathrm{NL}}$ parameter, as allowed by Planck constraints, yields the maximum contribution of the non-linear terms to $P_{gg}(k,\eta)$, even when $b_{\mathrm{NL}} = 0$. In such case, our plots show that values of order $g_{\mathrm{NL}} \sim -10^5$ could be detected (or ruled out) in the planned all-sky surveys. We thus conclude that $g_{\mathrm{NL}}$ should not be ignored in the search for primordial non-Gaussianities imprinted in the galaxy power spectrum. 

Let us emphasise that we have not included relativistic contributions to the light propagation like redshift space distortions or gravitational lensing (see e.g.~Ref.~\cite{2018MNRAS.478.1341K}). Adding these effects may alter, or even erase the imprints here presented, as happens with linear order relativistic signatures \cite{2020JCAP...11..064G,2021arXiv210608857C}. However, the existence of a signature beyond the predicted uncertainties of future surveys is a novel feature that deserves further attention. We leave this task to future work. 
While our results do not account for the full relativistic effects in the observed galaxy distribution, we are confident to have incorporated contributions from PNG and relativistic non-linearities into the theoretical (source) galaxy power spectrum.

\acknowledgments 
We thank Obinna Umeh for useful discussions and an anonymous referee whose comments allowed us to significantly improve this paper. RMC acknowledges support of a
studentship funded by Queen Mary University of London and CONACyT
grant No. 661285. AP is a UK Research and Innovation Future Leaders Fellow, grant MR/S016066/1. JCH acknowledges sponsorship from CONACyT through grant CB-2016-282569, and FORDECYT-PRONACES project No. 304001/2020, as well as grant PAPIIT-UNAM IN107521 
(\textit{Sector Oscuro y Agujeros Negros Primordiales}). 
KAM is supported  by STFC grants ST/P000592/1 and ST/T000341/1. 

\bibliographystyle{JHEP.bst}
\bibliography{biblio}

\end{document}